# Color Routing and Beam Steering of Single-Molecule Emission with a Spherical Silicon Nanoantenna

*María Sanz-Paz[1,2,\*], Nicole Siegel[1], Guillermo Serrera[3], Javier González-Colsa[3], Fangjia Zhu[1], Karol Kołątaj[1], Minoru Fujii[4], Hiroshi Sugimoto[4], Pablo Albella[3], and Guillermo P. Acuna[1,5,\*]*

[1] Department of Physics, University of Fribourg, Chemin du Musée 3, Fribourg CH-1700, Switzerland.

[2] Sorbonne Université, CNRS, Institut des NanoSciences de Paris, INSP, F-75005 Paris, France.

[3] Group of Optics, Department of Applied Physics, University of Cantabria, 39005, Santander, Spain.

[4] Department of Electrical and Electronic Engineering, Graduate School of Engineering, Kobe University, Kobe 657-8501, Japan.

[5] Swiss National Center for Competence in Research (NCCR) Bio-inspired Materials, University of Fribourg, Chemin des Verdiers 4, CH-1700 Fribourg, Switzerland.

[\*] Corresponding Authors: maria.sanz-paz@insp.upmc.fr, guillermo.acuna@unifr.ch

Abstract

Single-photon emitters radiate as electric dipoles, which limits light collection efficiency and complicates integration into flat photonic devices. Developing nanophotonic structures capable of directing photon emission with tunable angular distributions in the visible spectrum has been pursued for applications ranging from integrated optical systems to discrimination of molecular species. To date, such directional control has been achieved using components whose overall footprint is larger than the emission wavelength and often rely on lossy plasmonic components. Here, we employ the DNA origami technique for deterministic nanoscale assembly, positioning single fluorophores in nanometric proximity to a single silicon spherical nanoparticle and demonstrate unidirectional emission with forward-to-backward intensity ratios up to ~7 dB. Furthermore, we show that a single silicon nanosphere antenna can function as a color router or a beam steerer depending on its size, emitter spectral range and emitter-nanoparticle distance.

Introduction

Highly directional single-photon emitters are crucial for building efficient integrated photonic circuits and quantum communication systems.[1–3] Precise control over photon emission direction—through beam steering— or spectral (i.e., "color") routing, where different wavelengths are directed into opposing directions,[4] enables advanced applications in high-resolution color imaging,[5] hyperspectral imaging,[6–8] and optical spectroscopy.[9–12] Plasmonic nanoantennas have been widely investigated to tailor radiation patterns using grating structures[4] or Yagi-Uda-inspired designs,[13] among others. These devices achieved directionality through interference between multiple resonators, which typically requires dimensions exceeding $\lambda^2$ (with $\lambda$ denoting the emission wavelength). More compact configurations have later emerged, utilizing as few as two metallic components.[14,15]

Directionality can also result from interference between distinct optical modes. As first demonstrated by Kerker,[16] tuning the relative contribution of electric and magnetic dipolar modes can lead to the suppression of either backward or forward scattering (known as first and second Kerker conditions, respectively).[16,17] This approach was recently explored to achieve unidirectional emission using a self-assembled gold nanosphere trimer, where both the magnetic and electric dipolar resonances were excited by a single dipole emitter.[18]

Although metal-based nanoantennas can deliver high directivity with forward-to-backward ratios (F/B) exceeding 10 dB,[15] their performance is often narrow band[15,18] and sensitive to small variations in element size or emitter position.[15] These constraints hider their applicability in color routing or beam steering. Moreover, their radiation efficiency is significantly compromised by intrinsic optical losses due to metal absorption.

On the other hand, High Refractive-Index Dielectric (HRID) nanoparticles offer a low-loss alternative that supports strong electric and magnetic Mie resonances and flexible radiation-pattern engineering in the visible range.[19,20] The first experimental demonstration of directional scattering using HRID materials involved individual silicon nanoparticles (SiNPs),[19] followed by other designs such as notch SiNPs.[21] Dimers and lithography-designed V-shape antennas,[22,23] have also been proposed to enhance scattering directionality and achieve color routing of the scattered light.[22,23] Plasmonic designs like the Yagi-Uda geometry have been extended to dielectric platforms.[24,25] HRID metasurfaces for directional emission have been experimentally demonstrated,[26,27] including color routing of lanthanide emitters that allowed the spatial separation of electric and magnetic dipole emissions.[28]

Building on this foundation, numerical studies indicate that a single HRID NP can effectively direct emission from a nearby dipolar emitter,[24,29] drastically reducing the footprint of previous demonstrations. Experiments combining individual SiNPs with monolayers of transition metal dichalcogenides have demonstrated emission directionality at the ensemble level.[30,31]

Here, we experimentally demonstrate that a single SiNP with a footprint < 0.05 $\lambda^2$ can serve not only as an optical antenna capable of directing light from a nearby emitter, but also as a versatile element for color routing and beam-steering. Using DNA origami assembly,[32] we position single emitters adjacent to one SiNP with nanometric accuracy and map the resulting far-field emission patterns.[33] Using a 120 nm SiNP, we achieve robust broadband directionality in the visible range, with F/B values as high as 7 dB, for emitter-NP separations ranging from 8 to 85 nm. Furthermore, the emission direction follows the displacement of the emitter around the NP, realizing a beam-steering effect previously predicted theoretically.[34] For 170 nm SiNPs and specific emitter positions, the emission direction becomes wavelength-dependent, routing green and red emission into nearly opposite directions. These results are in good agreement with previous theoretical works. Moreover, specific simulations of the experiments elucidate the underlying modal interference mechanisms responsible for directionality, beam steering, and color routing, opening the way for further designs of HRID nanophotonic directional devices.

Sample fabrication

Fig. 1A illustrates the DNA origami used, consisting of a two-layer 12-helix rectangular structure with approximate dimensions of 180 nm × 20 nm × 5 nm (length × width × height) and a "mast" feature in the center[35] (detailed design parameter are provided in Tables S1 and S2). The design incorporates 10 biotinylated modifications on the bottom side to enable immobilization on glass coverslips functionalized with neutravidin. Additionally, it features 32 single-stranded (ss) DNA handles, composed of a mixture of $A_8$ and $A_{18}$ to accommodate a single SiNP at the upper side.[33] Finally, ATTO 542 (green) and/or ATTO 647N (red) fluorescent molecules can be incorporated at different positions (denoted $p_i$ with *i = -2, 1, 2, 3* in Fig. 1A). At each position $p_i$, one or two fluorophores can be hosted in close proximity, forming a pair.

Spherical SiNPs[36] were incorporated into DNA origami structures using a "surface synthesis" strategy.[37] In this approach, DNA origami templates were first immobilized on glass coverslips via biotin-neutravidin interactions. Subsequently, a solution containing DNA-functionalized

SiNPs was introduced (Fig. 1B) and incubated, followed by rinsing to remove the unbound NPs. We selected this approach over the "solution synthesis"[38] as it allows precise control over the hybridization efficiency between DNA origami structures and SiNPs by adjusting nanoparticle concentration and incubation time. This method provides an internal reference population (DNA origami structures without SiNPs), which can later be distinguished via colocalization analysis using scanning electron microscopy (SEM).[33] The resulting distance between each fluorophore and the SiNP surface, $d_i$, corresponding to each position, $p_i$, depends additionally on the NP size (a detailed calculation of $d_i$ can be found in Table S3). Importantly, we have recently characterized the photophysical interaction of fluorophores with SiNPs, finding a significantly reduced quenching even at short separations compared to metallic NPs,[33] which emphasizes another advantage of using HRIDs.

The emission directionality of the assembled nanoantennas was experimentally characterized using a custom-built scanning fluorescence microscope, modified to project the back focal plane (BFP) image of the structures (Fig. 1C) onto a CCD camera (see experimental methods for details). To ensure accuracy, only those structures confirmed by SEM to host a single SiNP were included in the analysis. Structures without nanoparticles served as reference controls. The F/B emission ratio was calculated from the BFP images using the so-called areal method.[15,39] It is obtained by integrating the fluorescence intensity over an azimuthal angular range of ±90° around the direction of maximum emission (defined as $\varphi$), and dividing it by the integrated intensity from the corresponding opposite direction (see highlighted sectors in Fig. 1C).

The crystalline and monodispersed colloidal SiNPs employed had diameters of either 120 or 170 nm. For these sizes, SiNPs in water exhibit Mie resonances within the visible spectrum, as seen in the numerical simulations of the scattering cross sections in Fig. 1D, which overlap differently with the two fluorophores used.

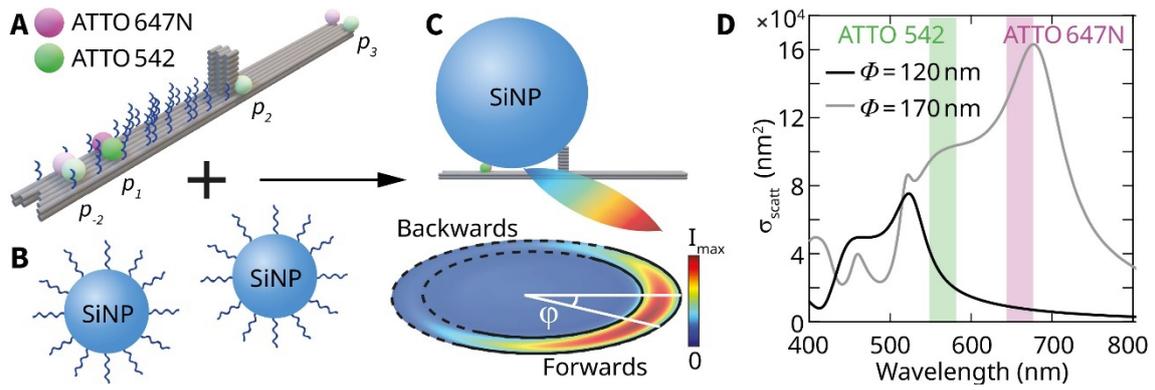

*Figure 1. Directional DNA origami-templated antenna.* (A) Sketch of the DNA origami used, displaying the different dye-NP distances $p_i$ (magenta and green dots) we studied and the polyadenine (blue protrusions) strands to attach subsequently added DNA-functionalized SiNPs (B). (C) Directional emission is characterized in the Fourier space by relaying the BFP image. From that image, the forward (solid line) and backward (dashed line) regions are used for forward-to-backward (F/B) quantification. The angle of maximum intensity, $\varphi$, is also indicated. (D) Simulated scattering cross-section for a SiNP of the diameters $\Phi$ used in this work in water. The shaded areas highlight the spectral emission bands of the two dyes used in our experiments (ATTO 542 in green and ATTO 647N in magenta).

Unidirectional Emission

To evaluate the capability of individual SiNPs to function as unidirectional optical antennas, we first employed NPs with a diameter of 120 nm combined with three distinct DNA origami structures differing solely on the position of the fluorophores at $p_1$, $p_2$ and $p_3$, (see Fig. 2A). Representative BFP images for each case are shown in Fig. 2B, with additional examples provided in Fig. S1. For all three positions, we observed directional emission from both the red and green fluorophores, as schematically illustrated in Fig. 2C. The extracted F/B emission ratios for ATTO 542 and ATTO 647N are presented in Fig. 2D and Fig. S2, respectively. These results confirm that the 120 nm SiNP is able to direct molecular emission across a broad frequency range and almost independently of the emitter–nanoparticle distance, in agreement with previous predictions.[30] Importantly, unidirectional emission was achieved despite the lack of control over the fluorophore dipole orientation (this will be later addressed in Fig. 6), further underscoring the robustness of this system. In contrast, control samples without NP (Fig. S3) exhibited no directional emission. ATTO 647N (first row in Fig. S3) typically produced a two-lobed pattern, indicating a fixed dipole orientation that results from a stable interaction of the fluorophore with the DNA scaffold. Conversely, ATTO 542 (second row in Fig. S3) displayed a circular emission pattern, characteristic of a freely rotating dipole—effectively an angular average of a dipolar emission profile. These results are consistent with earlier studies examining fluorophore orientation in DNA origami constructs across various configurations.[40,41]

We further characterized the performance of single SiNP antennas in terms of the azimuthal directivity imparted. Specifically, we extracted the angle of maximum emission $\varphi$. Due to the random orientation of DNA origami structures immobilized on glass coverslips, the absolute value of $\varphi$ is not relevant. Instead, we computed the differential emission angle between the red ($\varphi_r$) and green ($\varphi_g$) spectral channels ($\Delta\varphi = \varphi_r - \varphi_g$) for the same nanoparticle–fluorophore pair configuration, thereby revealing spectral dependencies in the angular emission profile (see Methods section for details on the quantification). It is worth mentioning that these studies are made possible by the DNA origami technique, which ensures the precise placement of a single fluorophore pair at a controlled distance from an individual SiNP. The distributions of $\Delta\varphi$ are spread with median values of -2º, 2º and -3º for *p₁, p₂* and *p₃,* respectively (Fig. 2E). These small-angle differences indicate that the emission of both fluorophores is directed by the SiNP essentially in the same direction, for all fluorophore-NP separations investigated, as also observed on the exemplary images in Fig. 2B. Finally, we note that the directivity imposed by the SiNP is quite robust against variations in nanoparticle size, as we observed a comparable behavior for 140 nm SiNPs (Fig. S4), and in fluorophore distance (Fig. 2D). Thus, we cannot attribute the spread in $\Delta\varphi$ observed in Fig. 2E to variations in those parameters, but rather to the accuracy of our analysis to determine the angle of maximum emission. Moreover, we have demonstrated the broadband performance of this antenna design within the visible range, in contrast to previously reported antennas.[13,15,42]

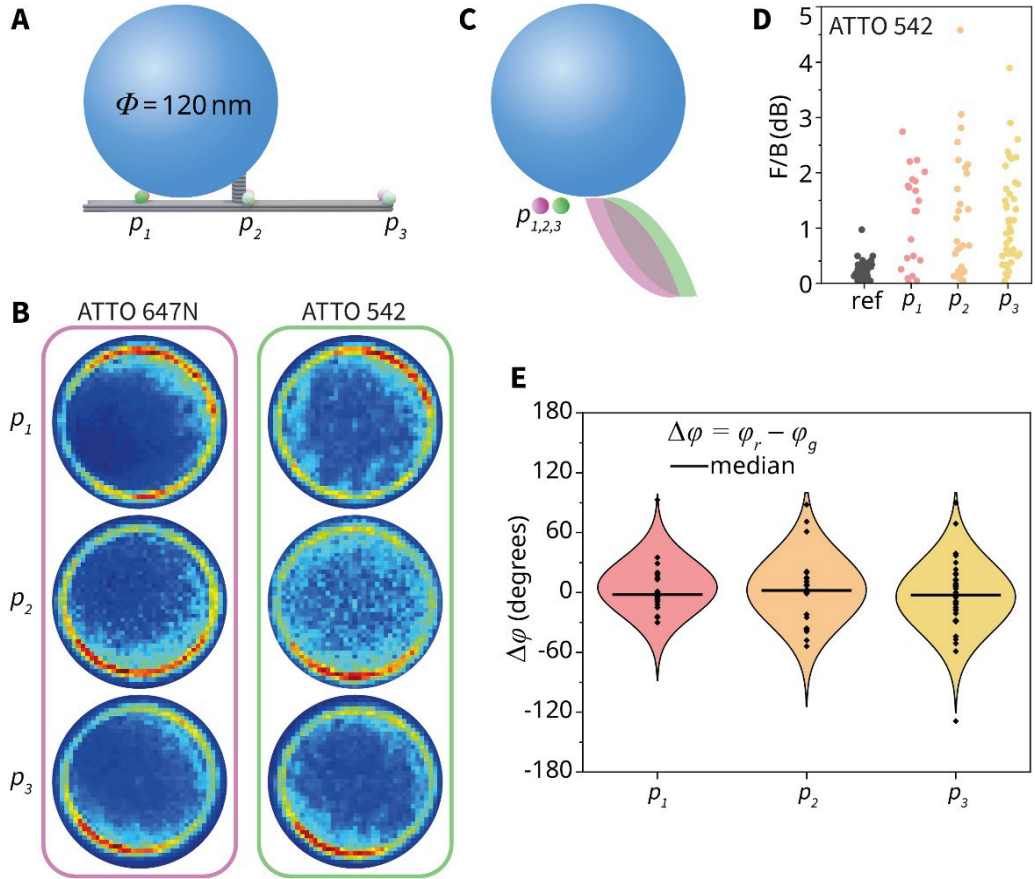

*Figure 2. Broadband directional emission with a single SiNP. (A) Illustration of the assembled monomer antennas for the three different dye-NP distances used ($p_1$, $p_2$ and $p_3$). (B) Exemplary BFP images for ATTO 647N (left) and ATTO 542 (right) from the same nanoantenna, shown according to the NP-emitter distance in (A). (C) Illustration of the nanoantenna behavior directing light from both fluorophores into the same direction for every distance. (D) Bee swarm distribution plots of the forward-to-backward (F/B) ratios obtained from the BFP patterns of ATTO 542 alone (ref) and at different distances $p_i$ to a SiNP. (E) Difference in the angle of maximum intensity, $\Delta\varphi$, between the green (ATTO 542) and red (ATTO 647N) spectral ranges for all three distances $p_i$. Dots correspond to individual data points, and envelopes represent a normal distribution. Median values: -2°, 2° and -3° for $p_1$, $p_2$ and $p_3$, respectively. Number of points: $N_1 = 18$, $N_2 = 23$ and $N_3 = 40$.*

Beam Steering

Building on the observed broadband directivity, we explored the possibility of dynamic control of the direction of the antenna's radiation pattern, the so-called beam steering,[34] which could further expand the functionality of this system. To this end, we positioned two fluorescent molecules (either ATTO 542 or ATTO 647N) at symmetric locations on opposite sides of a 120 nm SiNP (positions $p_{-2}$ and $p_2$ in Fig. 1A) resulting in similar absolute distances of ~20 nm (Fig. 3A). It is worth mentioning that the high degree of symmetry and simplicity of the SiNP antenna implies that lateral displacements of the dipole source induce a rotation in the radiation pattern[34] (illustration in Fig. 3B). Following the procedure from the previous experiments, a series of BFP images were acquired for each structure, from which intensity transients were generated. As shown in Fig. 3C, these intensity traces typically exhibit a two-step photobleaching behavior, with each step corresponding to the bleaching of one of the fluorophores. This separation in time enables isolation of time intervals corresponding to either dual or single-fluorophore emission. By subtracting the BFP images corresponding to these intervals, we extracted the individual directional emission patterns of each fluorophore. This approach enables a clear discrimination between the emission profiles of the two emitters (fluorophore-1 and fluorophore-2), as shown in Fig. 3D (additional examples are provided in Fig. S5). We further quantify the beam steering performance of SiNPs by extracting the differential angular emission, $\Delta\varphi = \varphi_1 - \varphi_2$ between the two fluorophores in the green and red spectral ranges studied. The results, included in Fig. 3E, are characterized by $\Delta\varphi$ median values of 156º and 160º amid a considerable spread. We attribute this spread in $\Delta\varphi$ values to the fixed dipole orientation in some cases (see ATTO 647N example in Fig. S3). The $\Delta\varphi$ median values confirm that a lateral shift in fluorophore position can reverse the emission direction. This finding demonstrates that single-molecule beam steering can be achieved simply by adjusting the emitter's position relative to the SiNP (as illustrated in Fig. 3B).

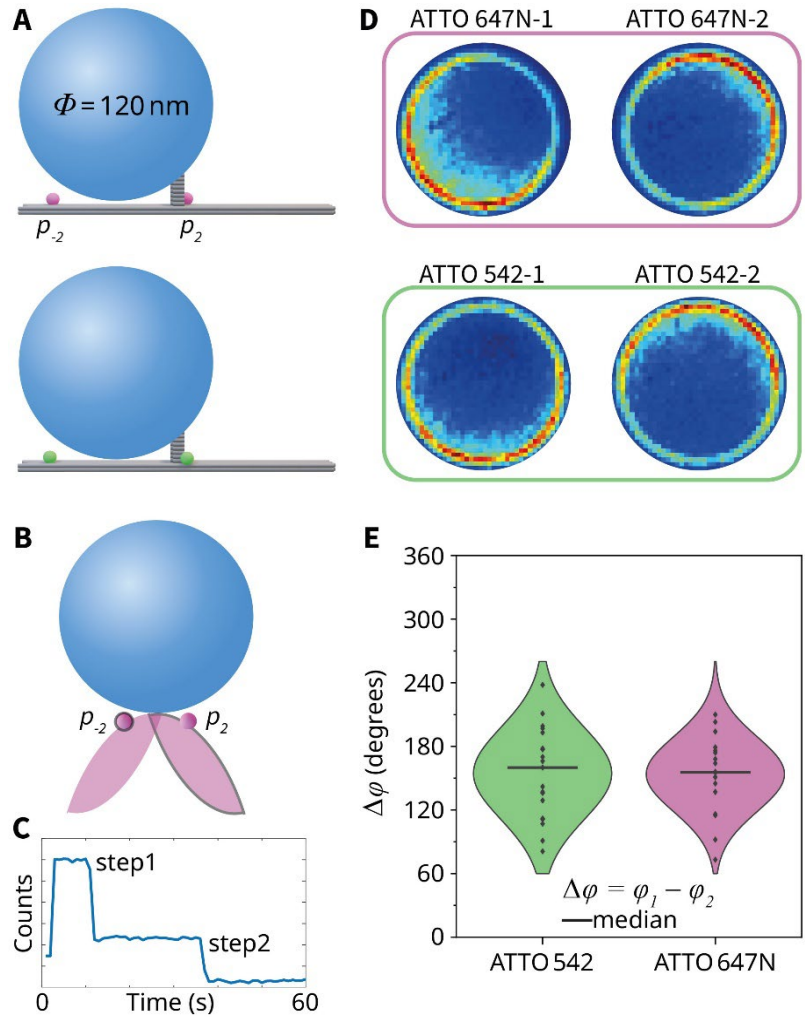

*Figure 3. Beam steering with a single SiNP.* *(A) Illustration of the assembled monomer antennas with the same fluorophore at equal distance and opposite sides ($p_2$ and $p_{-2}$) of the SiNP. (B) Exemplary BFP images of the two ATTO 647N (top) and ATTO 542 (bottom) in the same nanoantenna. (C) Illustration of the nanoantenna with two emitters (with and without contour) positioned at equal distances but on opposite sides, resulting in a shift in the emission direction (beam steering behavior). (D) Exemplary fluorescence time trace displaying two distinct bleaching steps, attributed to both dyes being positioned within diffraction-limited distances. (E) Difference in the angle of maximum intensity, $\Delta\varphi$, for two ATTO 542 (green) or two ATTO 647N (red) fluorophores positioned on opposite sides of the NP at the same distance. In average, the two fluorophores contained in the same structure emit light into opposing directions. Median values: 156° and 160° for ATTO 647N and ATTO 542, respectively. Number of points: $N_r = 20$ and $N_g = 19$.*

Color Routing

A further aspect of directivity control is spectral (or color) routing, whereby different wavelengths are guided along distinct spatial directions.[24,25] This mechanism is strongly dependent on NP size and NP-emitter distance, as these parameters affect the amplitude and relative phase of the electric and magnetic dipolar resonances excited in the NP. The numerical simulations shown in Fig. 4 illustrate this. A SiNP with a diameter of 120 nm does not exhibit color routing in the spectral regions corresponding to the emission of ATTO 542 and ATTO 647N, regardless of the emitter's position (Fig. 4A). In fact, for wavelengths >525 nm the F/B values remain positive for the three positions. This observation is confirmed by the results shown in Fig. 2E. By contrast, when the SiNP diameter is increased to 170 nm (Fig. 4B), a clear reversal in emission direction is observed between the emissions of ATTO 647N and ATTO 542 for an emitter placed at position $p_3$. However, this color routing vanishes for position distances $p_1$ and $p_2$. In addition, from the results in Fig. 4, we can also conclude that an increase in both NP diameter and emitter–NP distance leads to a redshift in the emission directivity, highlighting the versatility of this configuration for directional modulation.[31] In agreement with previous reports,[30] these simulations show that the wavelength corresponding to maximum F/B ratio is ~4 times the diameter, emphasizing the compact footprint required to achieve directional emission.

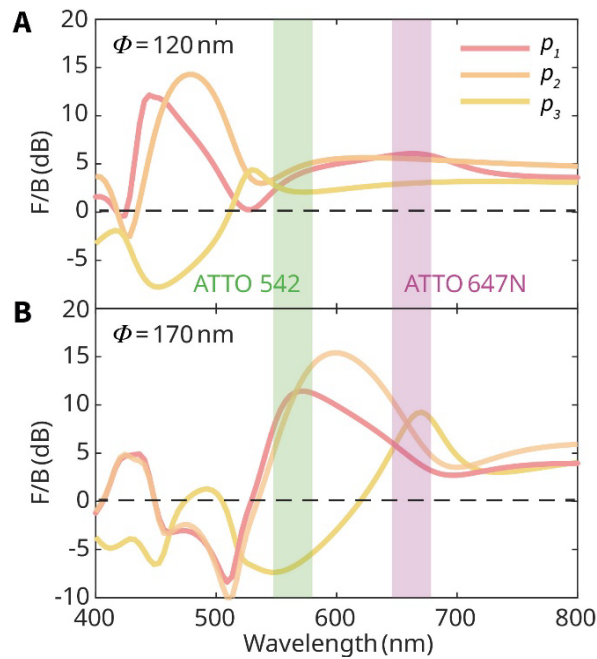

*Figure 4. Expected F/B as a function of the SiNP size. Numerical F/B calculations for a SiNP of diameter 120 nm (A) and 170 nm (B) excited by a dipolar emitter (average orientation)*

*located at positions p₁ (gray), p₂ (blue) and p₃ (green) from the SiNP surface. The shaded areas highlight the spectral emission bands of the two dyes used in our experiments (ATTO 542 in green and ATTO 647N in magenta). The dotted line marks the shift from forward (F/B>0) to backward (F/B<0) emission (i.e., to and away from the SiNP).*

Guided by the simulations presented in Fig. 4, we conducted measurements analogous to those shown in Fig. 2 using larger SiNPs with a diameter of 170 nm (Fig. 5A). Contrary to the behavior observed for smaller particles, where both types of fluorophores exhibited similar angular distributions regardless of their position, the larger SiNP clearly separates the emission of the fluorophores when they are placed at the largest emitter–NP separation, $p_3$ (Fig. 5B and Fig. S6). This behavior is consistent with simulations in Fig. 4B. A quantitative analysis of the angular separation computing $\Delta\varphi = \varphi_{\text{red}} - \varphi_{green}$ (Fig. 5E) revealed minor differences at the two closer positions ($p_1$ and $p_2$) and nearly opposite emission directions with a median $\Delta\varphi = 166°$ for position $p_3$. These results demonstrate that a single SiNP (see illustration in Fig. 5D) can spatially separate the emission of two spectrally different fluorophores in close proximity, effectively enabling color routing at the single molecule level. It is worth noticing here that, according to the simulations in Fig. 4B, we are now in a range where the direction of the emission is more sensitive to changes in both SiNP diameter and fluorophore position. This explains the larger spread in $\Delta\varphi$ observed in Fig. 5E, specially for the case of $p_3$, where even some cases are observed showing opposing behavior (see examples in Fig. S6).

The experimentally obtained F/B ratios across multiple SiNP–fluorophore configurations (Fig. S7) exhibit enhanced directionality for the larger SiNPs compared to the 120 nm case (Fig. 2D and Fig. S2). A modest reduction in directionality is observed with increasing emitter–NP distance, consistent with simulations in Fig. 4. Note that simulated F/B values correspond to a specific wavelength, so experimental ones appear to be slightly lower since they come from a spectral range.

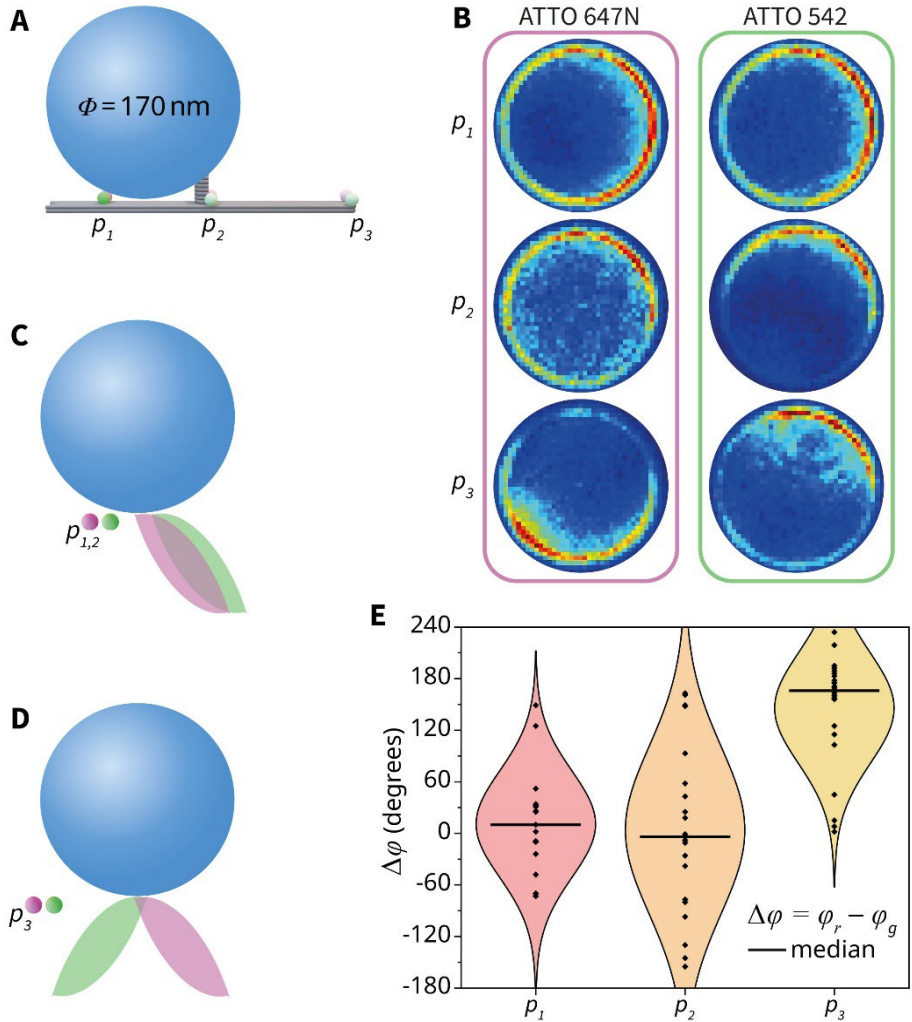

*Figure 5. Color routing.* (A) Illustration of the assembled monomer antennas for the three different dye-NP distances used ($p_1$, $p_2$ and $p_3$). (B) Exemplary BFP images for ATTO 647N (left) and ATTO 542 (right) from the same nanoantenna, shown according to the NP-emitter distance in (A). (C) Illustration of the nanoantenna behavior directing light from both fluorophores into the same direction for $p_1$ and $p_2$. (D) Illustration of the color routing nanoantenna behavior for $p_3$. (E) Difference in the angle of maximum intensity, $\Delta\varphi$, between the green (ATTO 542) and red (ATTO 647N) spectral ranges for all three distances $p_i$. A clear shift is seen when increasing the distance. Median values: 10°, -4° and 166° for $p_1$, $p_2$ and $p_3$, respectively. Number of points: $N_1 = 17$, $N_2 = 22$ and $N_3 = 25$.

Interestingly, the observed distance dependency of color routing is consistent with a transition in behavior for the green emission observed in our simulations as well as in previous theoretical predictions by Rolly *et al.*[29,43] As the emitter–NP separation increases, the system transitions from a collector regime where the NP guides energy forward, (Fig. 5C) to a reflector regime where the emission is redirected backward (Fig. 5D). This distance-dependent transition between collector and reflector regimes can be interpreted in terms of induced modes in the emitter-NP system, which give rise to generalized Kerker conditions where radiation in one direction is suppressed. To elucidate the nature of these induced multipoles, we analytically calculated the far field radiation of the emitter-NP system embedded in a homogeneous medium with refractive index $n$ = 1.33, following the theoretical framework originally developed by Kerker and coworkers.[44,45] Polar radiation plots representing radiation in the *XZ* plane were calculated as shown in Fig. 6A. The dipolar emitter ***p*** was located below and to the right with respect to the SiNP. In this geometry, angles within the [270º, 90º] interval correspond to backward (reflected) radiation, whereas angles within the [90º, 270º] interval represent forward (collected) radiation. The shadowed green areas ([180º, 225º] and [315º, 360º] intervals) in Fig. 6A represent the integration areas of BFP images (as in Fig. 4).

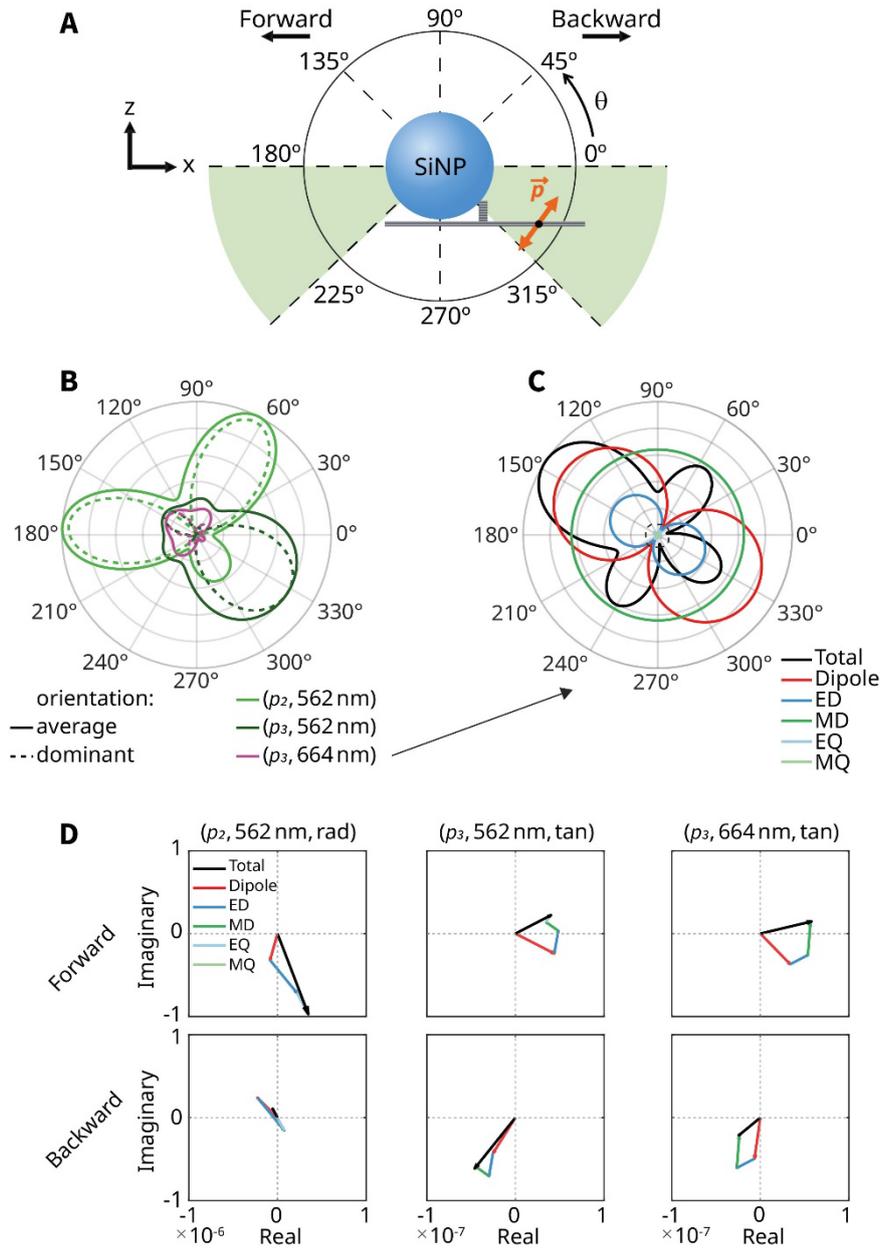

*Figure 6. Multipolar interpretation. (A) Schematic of the geometrical framework used for the analytical calculations. (B) Solid lines: emitter orientation-averaged radiation patterns in the XZ plane for representative cases at different distances and emission wavelengths ($p_2$, 562 nm; $p_3$, 562 nm and $p_3$, 664 nm), illustrating the color-routing effect at $p_3$ and the distance-dependent change in direction at 562 nm. The wavelengths 562 nm and 664 nm correspond to the peak emissions of ATTO 542 and ATTO 647N, respectively. Dashed lines: radiation patterns for the dominant emitter orientation in each case: radial to the NP for $p_2$ and tangential for $p_3$. (C) Multipolar decomposition of the far-field emission for the case ($p_3$, 664 nm) with the emitter tangential to the NP. The red curve indicates the direct emitter contribution to the overall emission, while the patterns labeled as ED (electric dipole), MD (magnetic dipole), EQ (electric quadrupole) and MQ (magnetic quadrupole) correspond to the induced multipoles in the SiNP.*

*(D) Phasor representation of the multipolar contributions to the far-field for the dominant emitter orientation in the situations shown in (B), i.e., radial for $p_2$ and tangential for $p_3$.*

Fig. 6B shows emitter orientation-averaged radiation patterns in the *XZ* plane for representative situations at different distances $p_i$ and emission wavelengths measured with the 170 nm SiNP (as in Fig. 4). Since the near-field of a dipolar emitter is stronger along its oscillation axis, emitters radially oriented (or close to radial) with respect to the NP provide the dominant contribution to the averaged radiation pattern when positioned near the NP ($p_1$ and $p_2$). This explains the ($p_2$, 562 nm) case in Fig. 6B, where the total far field is highly directional in the forward direction with respect to the position of the emitter (in agreement with Fig. 4B), and features two "oblique" lobes (180º and 60º, approximately), with a smaller central lobe in the backward direction (~300º). The dashed green line, which corresponds to the contribution from that radial orientation, dominates the overall pattern and causes the lateral oblique lobes. Since the excitation fields are radial to the sphere, only electric resonances are excited, resulting in the suppression of radiation in either the forward or backward directions depending on the wavelength. The residual backward lobe arises from contributions of other emitter orientations with smaller amplitude. Therefore, the strong forward directionality observed under similar situations in Fig. 4B can be attributed primarily to the excitation of radial electric resonances by a radially orientated dipole.

The opposite situation occurs for larger distances ($p_3$), where the radiation of the dipolar emitter along its axis is suppressed. Thus, the dominant contributions will come from emitters oriented tangential to the NP, and radiation patterns will feature stronger radiation in the forward and backward directions depending on the wavelength (~150º for 664 nm and ~300º for 562 nm, respectively), as shown in Fig. 6B. This is again supported by the corresponding dashed lines, which here correspond to a tangential orientation, that matches the forward and backward directions maxima. Particularly, the two analyzed cases, corresponding to the two different emission wavelengths (562 and 664 nm), feature a reversal of the radiation direction. This is the color-routing effect observed experimentally (Fig. 5E). Moreover, comparison between $p_2$ and $p_3$ at 562 nm also shows an inversion of the direction, further supporting the above-mentioned change from a collector to a reflector behavior when increasing the NP-emitter distance. Overall, the existence of a dominant dipolar orientation in every case is the reason behind our system's robustness to the lack of control over the fluorophore orientation.

To visualize the multipolar components at play, we plot in Fig. 6C the far field decomposition corresponding to case ($p_3$, 664 nm) for an emitter oriented tangential to the NP (dominant one). It can be seen that the main contributions come from emitter radiation and the ED and MD induced modes, with small contributions from EQ and MQ, depicted in the inset. Additional results for other situations with separate contributions from individual orientations can be found in Fig. S8. We must emphasize the role of the substrate in the experiments. Although its influence is expected to be minor due to the small refractive-index contrast with the aqueous medium, it can nonetheless modify the observed radiation patterns. For instance, dipolar emission in the presence of a substrate typically propagates at angles close to the interface rather than along the tangential direction. Consequently, a full quantitative comparison to the experimental data cannot be achieved using this analytical approach.

To understand the interplay between these contributions, Fig. 6D shows phasor representations of the far field in the forward and backward angles[30] for the dominant orientation in each of the situations in Fig. 6B. For the case ($p_2$, 562 nm), the observed forward directionality can be understood as constructive/destructive interference of ED and EQ resonances. On the other hand, the $p_3$ cases are a result of ED/MD interplays with the dipolar emitter, where backward radiation switches from a constructive interference at 562 nm to a destructive interference at 664 nm. This switch, akin to a generalized Kerker condition, yields the color-routing behavior observed experimentally. More details on these results, including definitions of forward and backward radiation, can be found in the Supplementary Information.

Conclusions

In summary, our findings represent the experimental demonstration of directional emission by coupling a single dipole emitter with a single dielectric nanoparticle. The simplicity of this design offers a robust and adaptable platform for achieving unidirectional emission, with minimal sensitivity to nanoparticle size, emitter-nanoparticle distance, or orientation. More complex designs, such as notched dielectric NPs,[21,34] moderate-index dielectrics (which exhibit broader overlapping MD and ED resonances),[46] and hybrid metal-dielectric structures[47–49] have been suggested as promising alternatives to further boost directionality.

Moreover, the broadband behavior observed in this work paves the way for unidirectional white light,[50] previously unimaginable with structures like plasmonic Yagi-Uda designs that yield unidirectional emission in a narrow wavelength range. Importantly, this work also marks the first successful demonstration of fluorescence color routing using dielectric nanoparticles, with minimal spatial crosstalk, as the spatial separation is approximately 180° between the red and green spectral ranges.[22] Furthermore, by modifying the NP size and position, the spectral range of the directionality can be tuned,[24,25] and the NP behavior can be changed from collector to reflector.[29,43] Notably, routing was also achieved through lateral displacement of the emitter, i.e., beam steering,[34] a phenomenon never demonstrated experimentally. Our results are well explained in terms of the various induced multipoles in the emitter-NP system, where destructive/constructive interference between them depending on distance and emission wavelength gives rise to directional radiation.

Together, these findings represent crucial steps toward the development of compact, highly tunable directional antennas, which may find applications for on-chip nanophotonics and for sensing of multiple molecular species.

Methods

*DNA functionalization of SiNPs.* Crystalline SiNPs were synthesized following established protocols[36] and size-separated using a sucrose gradient centrifugation process. Initially, a gradient-forming system from Biocomp was employed to mix 50 wt% sucrose (placed at the bottom) and 20 wt% (placed at the top) using a SW40 rotor at 30 rpm, with a 12-second stop between rotations. The SiNP solution (0.5 wt% in methanol) was then carefully deposited on top, ensuring that the pipette tip made contact with the gradient surface to avoid breaking the surface tension. The samples were centrifuged at 2000 rcf for 90 minutes. Size-separated SiNP solutions were retrieved using an automated piston that gradually descended from top to bottom at a rate of 0.5 mm sec$^{-1}$, collecting fractions every 3 mm. The solution was washed by centrifugation and resuspension twice with water and twice with methanol. Subsequently, SiNPs 0.04 wt% (diameter = 120 or 170 nm with SD ~12 nm as calculated from the extinction simulated spectra[33]) were dispersed in anhydrous DMF (10 mL) by ultrasonication, followed by the addition of Chloro-propyl-trimethoxy silane (CPTMS) with a ratio of 200 CPTMS molecules per nm² of NPs. After 2 hours of sonication at 70°C, sodium azide was added in excess, and the mixture was stirred overnight at 40°C. The NPs were then purified by centrifugation at 2000 rcf for 10 minutes, followed by consecutive redispersion twice in methanol and twice in water.[51] Finally, freezing-assisted strain-promoted azide-alkyne cycloaddition was employed for DNA conjugation. Shortly, SiNPs (1 mL, 0.1 wt%) were centrifuged at 2000 rcf for 10 minutes and resuspended in a 3:1 mixture of DBCO-modified T18 and T36 (300 μL, 100 μM, Biomers GmbH). Subsequently, 1 mL of PBS-SDS-Tween20 buffer (1x Phosphate-Buffered Saline–137 mM NaCl, 2.7 mM KCl, 10 mM $Na_2HPO_4$, and 1.8 mM $KH_2PO_4$ – pH 7.5, 0.1% sodium dodecyl sulfate, 0.1 % Tween20) was added to the mixture, which was then frozen at -20°C for 2 hours. The NPs were thawed by sonication for 5 minutes, centrifuged twice for 10 minutes at 2000 rcf, and resuspended in water. For further purification and size separation, a 0.5 wt% agarose electrophoresis gel (LE Agarose, Biozym Scientific GmbH) in 0.5x TAE (20 mM Tris, 5 mM Acetate, 0.5 mM EDTA) and 6 mM $MgCl_2$ was run for 3 hours at 100 V. The NPs were recovered by extracting the desired bands from the gel.[52] The extinction spectra of the NPs were measured, and the size was calculated.

*DNA origami synthesis.* The DNA origami employed in this study was designed using CaDNAno,[53] and the structure files are available at https://nanobase.org/structure/146.[54] A 7249-nucleotide-long scaffold extracted from the M13mp18 bacteriophage (Tilibit Nanosystems GmbH) was folded into the desired shape using 243 staples in 1xTAE (40 mM Tris, 10 mM Acetate, 1 mM EDTA), 12 mM $MgCl_2$, pH 8 buffer. It was mixed in a 5-fold excess of staples (purchased from IDT) over scaffold, and 50-fold for the functional staples (fluorophores, biotin, and handles purchased from Biomers GmbH) shown in Tables S1 and S2. The mixture was heated to 70 °C and cooled down at a rate of 1 °C every 20 minutes up to 25 °C. The DNA origami structures were later purified by 1% agarose (LE Agarose, Biozym Scientific GmbH) gel electrophoresis at 70 V for 2 h and stored at -20 °C.[55]

*On-surface assembly of Optical Nanoantennas.* Glass slides with custom-made chromium grids were cleaned through a series of sonication baths, using the following solvents in sequence: 15 minutes in acetone, 15 minutes in isopropanol, 15 minutes in potassium hydroxide (3M), and 15 minutes in water. The surfaces were then activated by exposure to UV-ozone cleaning and subsequently coated with BSA-Biotin and Neutravidin (BSA-biotin and neutravidin, 0.5mg mL$^{-1}$ in PBS), each incubated on the grid for 25 minutes. The DNA origami (30 pM) was incubated on the surface for 15 minutes and attached via Biotin protrusions from the bottom of the structure. Finally, 2 pM of functionalized SiNPs in PBS-SDS-Tween20 buffer (1X PBS pH 7.5, 0.1 % SDS, 0.1 %Tween20) were incubated overnight.

*Optical measurements.* Measurements were performed on an inverted microscope (Olympus IX71). Excitation was performed with a randomly polarized supercontinuum white light laser (FYLA SCT1000) that was spectrally filtered to a wavelength of (635 ±5) nm or (532 ± 5) nm to efficiently excite the dye (ATTO 647N or ATTO 542, respectively). A high NA objective was used (Olympus, 100× NA=1.4) for excitation and collection. Emitted fluorescence is filtered with a combination of a long-pass and a band-pass filter (for ATTO 647N or ATTO 542) during signal collection to block laser excitation.

A flip mirror is incorporated at the exit of the microscope to switch between confocal and BFP imaging. For each region imaged, measurements were taken first using red excitation to ensure all ATTO 647N dyes are bleached before moving to green excitation in order to avoid FRET between both fluorophores. Confocal imaging is first employed to identify the location of the structures. Then, the detection is switched to BFP imaging, and the stage is moved to bring the targeted single structure to the excitation focal spot. Briefly, the BFP imaging path consists of a relay lens ($f$ = 100 mm) and a Bertrand lens ($f$ = 50 mm) that produce a BFP image onto an EM-CCD (Andor iXon Ultra 888). The confocal detection is performed with a single-photon counting APD (τ-SPAD, PicoQuant).

The EM-CCD is used in the kinetic mode to acquire videos. An integration time of 1 s per frame is applied, and a sufficient number of frames are recorded (~60 frames per nanoantenna) to allow for the observation of the bleaching step and to subsequently record BFP images of the emission. 4 × 4 binning is applied during acquisition to have a higher signal per pixel. To achieve a good signal-to-noise ratio, the EM gain of the camera is set to 75.

For BFP video analysis, an algorithm is used to determine the circular region of interest (ROI) containing the radiation pattern. The total signal within that region is extracted and plotted against time, producing an emission intensity time trace that identifies the bleaching moment. The frames before (fluorescence) and after (background) the bleaching step are averaged, and subtraction of these two time-averaged images yields the averaged background-free single-molecule fluorescence image that will be used for further analysis. Pixels with a peak intensity above a certain threshold (0.85 in most of our analysis) are divided into two groups based on their positions relative to the center of the ROI. The group showing the highest intensity will determine the forward direction, and the other one the backwards radiation. For the pixels in the forwards group, we determine the angle $\varphi$ of maximum emission as the central angle between the maximum and minimum angles in this group. The central angle for the backwards group of pixels is set to be 180 degrees away from that. This angle $\varphi$ was then used to compute the difference in emission directionality between two fluorophores as

$$\Delta\varphi = (\varphi_r - \varphi_g + 540) \bmod 360 - 180$$

for the case when they emit to the same direction, and

$$\Delta\varphi = |(\varphi_r - \varphi_g + 180) \bmod 360 - 180|$$

when they radiate into opposite angles.

Then, the mean distance of the pixels in each group from the center of the ROI is calculated. An inner and outer circle are defined with a radius equal to this mean distance minus or plus 2 pixels. The integral ratio of the pixel intensities bounded by these inner and outer radii on an angular range $\varphi$ of $\pm 90°$ around the maximum emission and on the opposite direction was used to quantify the forward-to-backward ratio (F/B), as previously described.[15]

*Forward to backward numerical calculations.* Commercially available software (Ansys Lumerical) is used to perform Finite-Difference Time-Domain (FDTD) simulations. A single silicon sphere of various diameters (120 and 170 nm) is placed on a glass (n = 1.51) substrate that extends through the lower infinite half-space. The distance between the particle and the substrate is 5 nm, and the NP center was taken as $x = 0$. A dipolar emitter is placed at the axial middle plane between the SiNP and the substrate, and located at three different lateral positions ($x = 25$, $x = 50$, and $x = 130$ nm) with respect to the SiNP center. A finer mesh of 1 nm is used in a region containing the SiNP and the dipole emitter. The BFP image of emission is obtained by projecting the electric field data of a large monitor placed close to the nanoantenna on a hemisphere of radius 1 m ('farfield' projection). This monitor is placed on the substrate side. Simulations were performed for the three dipole orientations and then used to obtain an orientation-averaged BFP pattern (see some examples in Fig. S9 for representative wavelengths). The F/B value at each wavelength (Fig. 4) is calculated by integrating the pixel intensities on the averaged BFP image for polar ranges of $\theta = 45° - 90°$ and $\varphi = 0° \pm 90°$ and $180° \pm 90°$ (to and away from the SiNP, respectively).

*Multipolar analytical calculation of far-field patterns.* The total field from the emitter-NP system was obtained as[44,45]

$$\boldsymbol{E}_{total} = \boldsymbol{E}_{em} + \boldsymbol{E}_{scat} \qquad (1)$$

The fields from the emitter and the NP scattering were obtained from the Mie-like series:[30]

$$\boldsymbol{E}_{em} = \sum_{\nu} D_{\nu}\left[s_{\nu}\boldsymbol{N}_{\nu}^{(3)} + t_{\nu}\boldsymbol{M}_{\nu}^{(3)}\right] \qquad (2a)$$

$$\boldsymbol{E}_{sca} = \sum_{\nu} D_{\nu}\left[u_{\nu}\boldsymbol{N}_{\nu}^{(3)} + v_{\nu}\boldsymbol{M}_{\nu}^{(3)}\right] \quad (2b)$$

where $\nu = l, m, \sigma$ ($l$ and $m$ are the multipole indices while $\sigma = e, o$ refers to the parity), $\boldsymbol{M}_{\nu}^{(3)}$ and $\boldsymbol{N}_{\nu}^{(3)}$ are the magnetic and electric Vector Spherical Harmonics[56,57] (where the (3) superindex denotes the spherical Hankel functions) and $s, t, u, v$ are multipolar coefficients. The dipole emitter is located at a distance $r_s = r + R$ above the NP with radius $R$, and is characterized by a dipolar moment $\boldsymbol{p}$, which can be tangential ($\boldsymbol{p} = p\,\hat{\boldsymbol{\theta}}$) or radial ($\boldsymbol{p} = p\,\hat{\boldsymbol{r}}$) to the direction of the NP center. For the tangential orientation, the expressions for $s, t, u$ and $v$ are:

$$s_{l1e} = \frac{p}{2\pi\varepsilon_b}\frac{ik^3}{x_s}\psi_l'(x_s) \quad (3a)$$

$$t_{l1o} = \frac{p}{2\pi\varepsilon_b}\frac{ik^3}{x_s}\psi_l(x_s) \quad (3b)$$

$$u_{l1e} = -a_l\frac{p}{2\pi\varepsilon_b}\frac{ik^3}{x_s}\xi_l'(x_s) \quad (3c)$$

$$v_{l1e} = -b_l\frac{p}{2\pi\varepsilon_b}\frac{ik^3}{x_s}\xi_l(x_s) \quad (3d)$$

with all other coefficients being zero and $D_{\nu} = \frac{2l+1}{2l(l+1)}$. $\varepsilon_b$ is the background medium permittivity, $k$ is the wavenumber and $x_s = kr_s$. $\psi_l$ and $\xi_l$ are the Riccati-Bessel and Riccati-Hankel functions, with $'$ denoting their derivatives with respect to their arguments. $a_l$ and $b_l$ are the well-known plane-wave Mie coefficients[56].

For the radial orientation of the emitter, the non-zero coefficients are given by:

$$s_{l0e} = \frac{p}{\pi\varepsilon_b}\frac{ik^3}{x_s}\psi_l(x_s) \quad (4a)$$

$$u_{l0e} = -a_l\frac{p}{\pi\varepsilon_b}\frac{ik^3}{x_s}\xi_l(x_s) \quad (4b)$$

with all other coefficients being zero and $D_{\nu} = \frac{2l+1}{2}$. Thus, the far field calculation consisted on the evaluation of equation 1 with the derived coefficient expressions for a given orientation at a large enough radial distance. For direct comparison with Lumerical FDTD, a dipole amplitude $p = 2.76 \cdot 10^{-31}$ Cm and a radial distance of 1 meter were chosen. The orientation-averaged radiation patterns in Fig. 6B were obtained by averaging the radiation patterns in the *XZ* plane for both orientations plus the *YZ* plane for the radial orientation (accounting for the azimuthal degeneracy in that orientation). For multipolar analysis the emitter part corresponds to the whole series in equation 2a, while the different scattering terms of the NP were extracted in equation 2b. The phasor analysis in Fig. 6D corresponds to the *θ* far field component, which was stronger for all analyzed cases.

**Acknowledgments**

G.P.A. acknowledges support from the Swiss National Science Foundation (IC00I0-228177), and the National Center of Competence in Research Bio-Inspired Materials NCCR (51NF40_182881).

# Supplementary Information

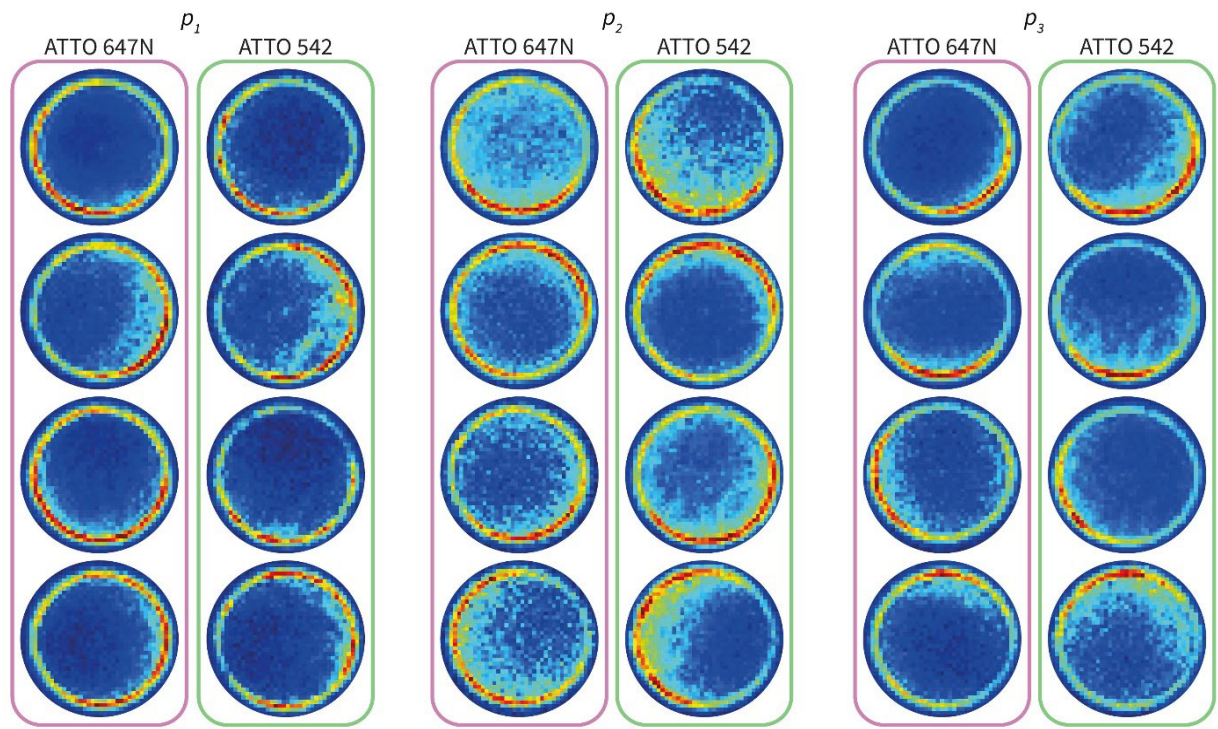

**Figure S1: Broadband directional emission with a single 120 nm SiNP.** Exemplary BFP images of ATTO 647N and ATTO 542 emission corresponding to the same structure, for the fluorophore pair positioned at distances $p_1$ (left), $p_2$ (middle) and $p_3$ (right) away from a 120 nm SiNP. Both fluorophores emit in the same direction.

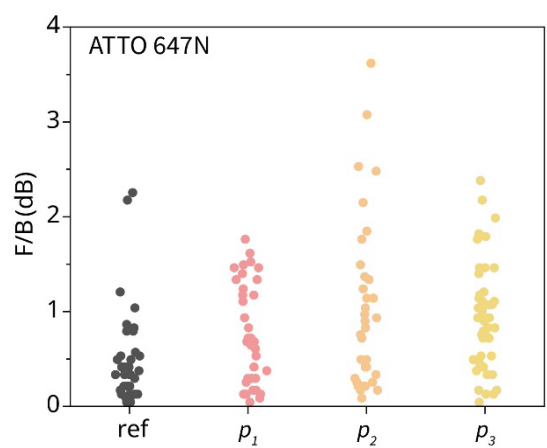

**Figure S2: F/B quantification for 120 nm SiNP.** Bee swarm distribution plots of the forward-to-backward (F/B) ratios obtained from the BFP patterns of ATTO 647N alone (ref) and at different distances $p_i$ to a 120 nm SiNP.

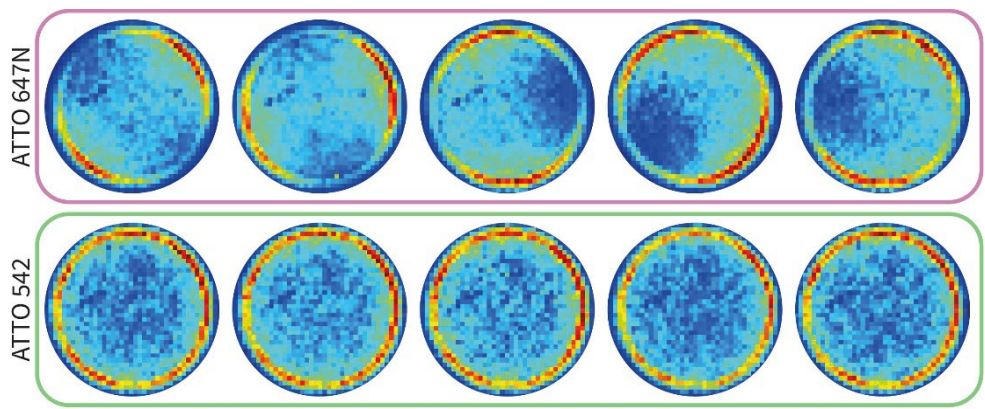

**Figure S3: BFP images in the absence of SiNPs.** Exemplary BFP images of ATTO 647N (top) and ATTO 542 (bottom) emission for origami structures in the absence of SiNPs (reference). ATTO 647N shows always a dipolar pattern, whereas ATTO 542 displays an isotropic emission characteristic of a rotating dipole.

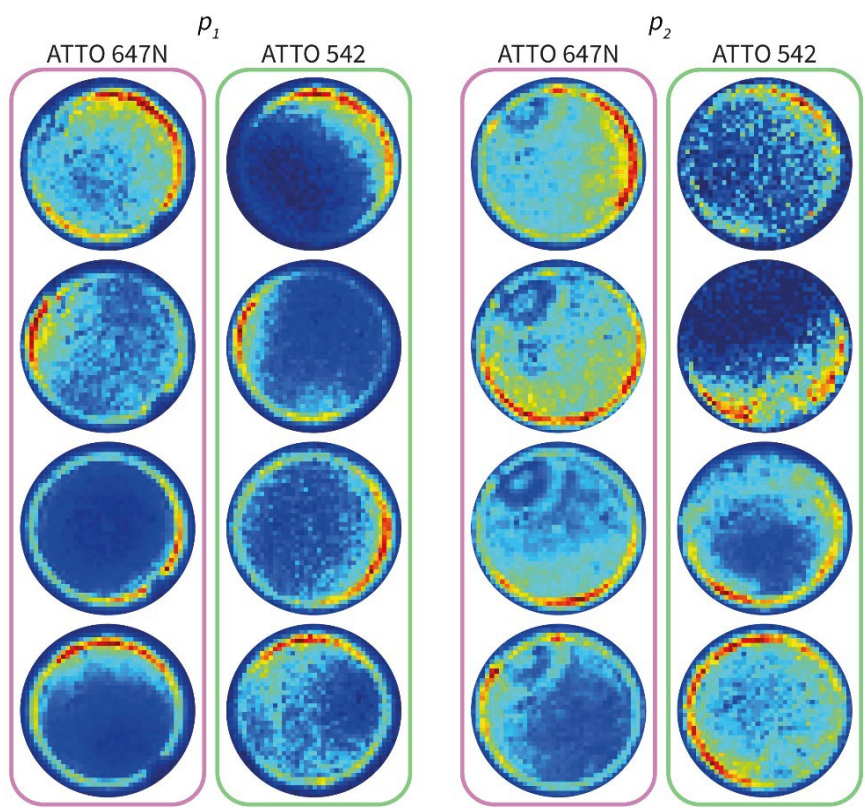

**Figure S4: Broadband directional emission with a single 140 nm SiNP.** Exemplary BFP images of ATTO 647N and ATTO 542 emission corresponding to the same structure, for the fluorophore pair positioned at distances $p_1$ (left) and $p_2$ (right) away from a 140 nm SiNP. Both fluorophores emit in the same direction.

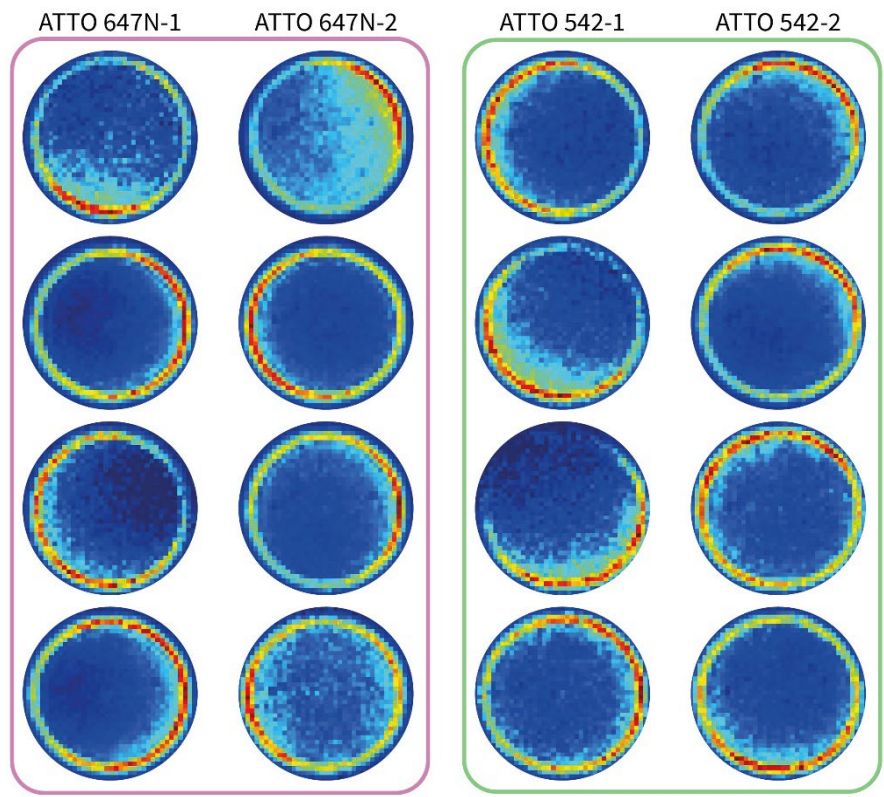

**Figure S5: Beam steering with a single SiNP.** Exemplary BFP images of two ATTO 647N (left) or two ATTO 542 (right) placed at opposite sides $p_{\pm 2}$ of a 120 nm SiNP. For each structure, the two fluorophores were labeled 1 and 2 according to their bleaching time. They emit into opposite directions.

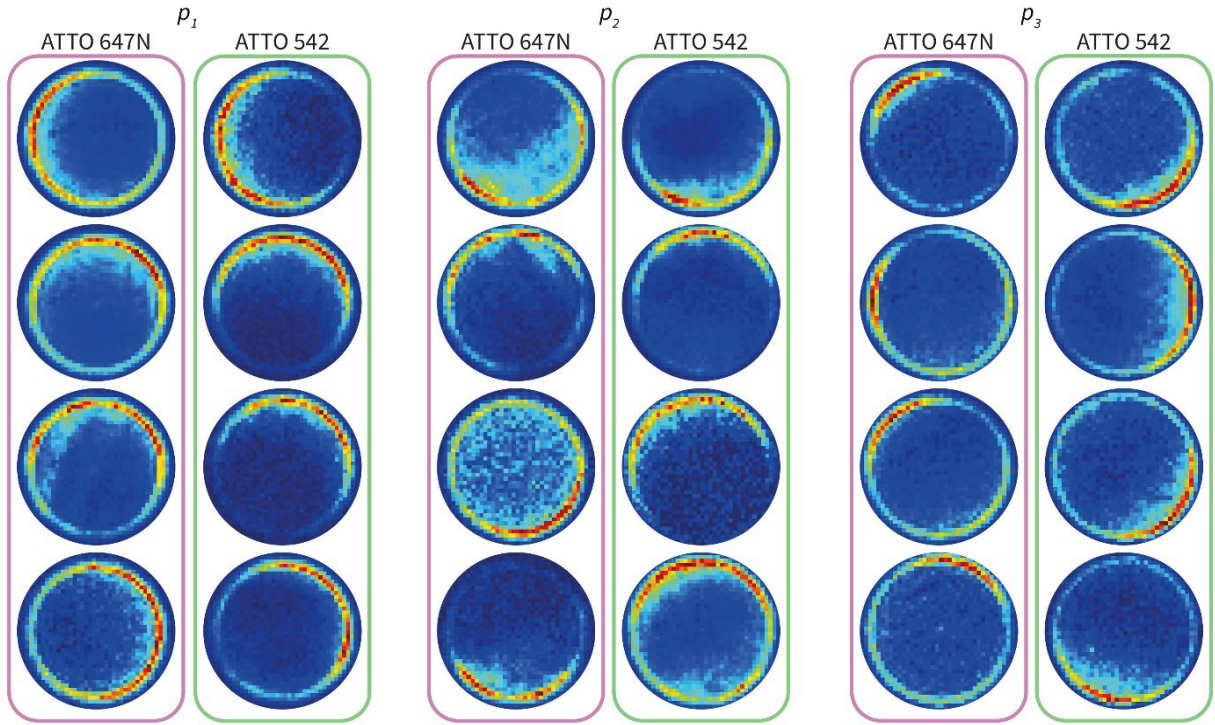

**Figure S6: Color Routing with a single 170 nm SiNP.** Exemplary BFP images of ATTO 647N and ATTO 542 emission corresponding to the same structure, for the fluorophore pair positioned at distances $p_1$ (left), $p_2$ (middle) and $p_3$ (right) away from a 170 nm SiNP. Color routing is observed for $p_3$.

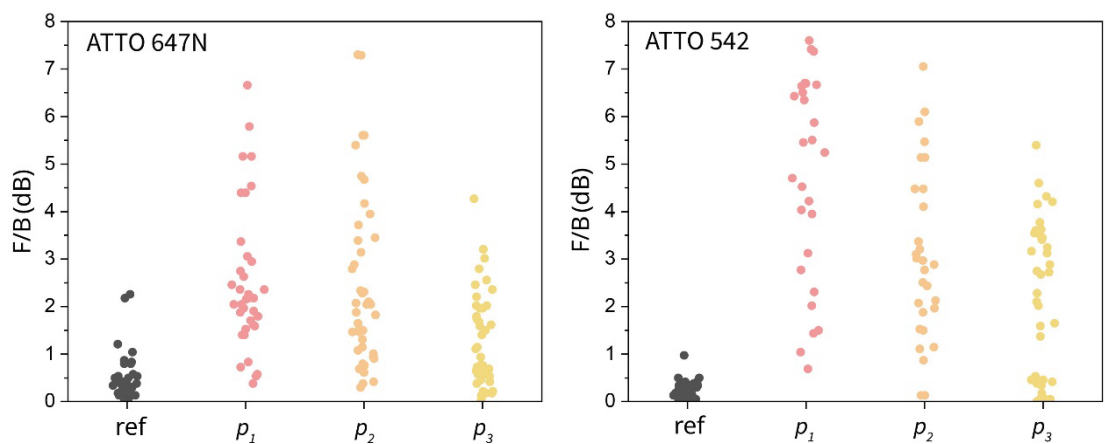

**Figure S7: F/B quantification for 170 nm SiNP.** Bee swarm distribution plots of the forward-to-backward (F/B) ratios obtained from the BFP patterns of ATTO 647N (left) and ATTO 542 (right) alone (ref) and at different distances $p_i$ to a 170 nm SiNP. Values are higher compared to those in Figure 2D and Figure S3.

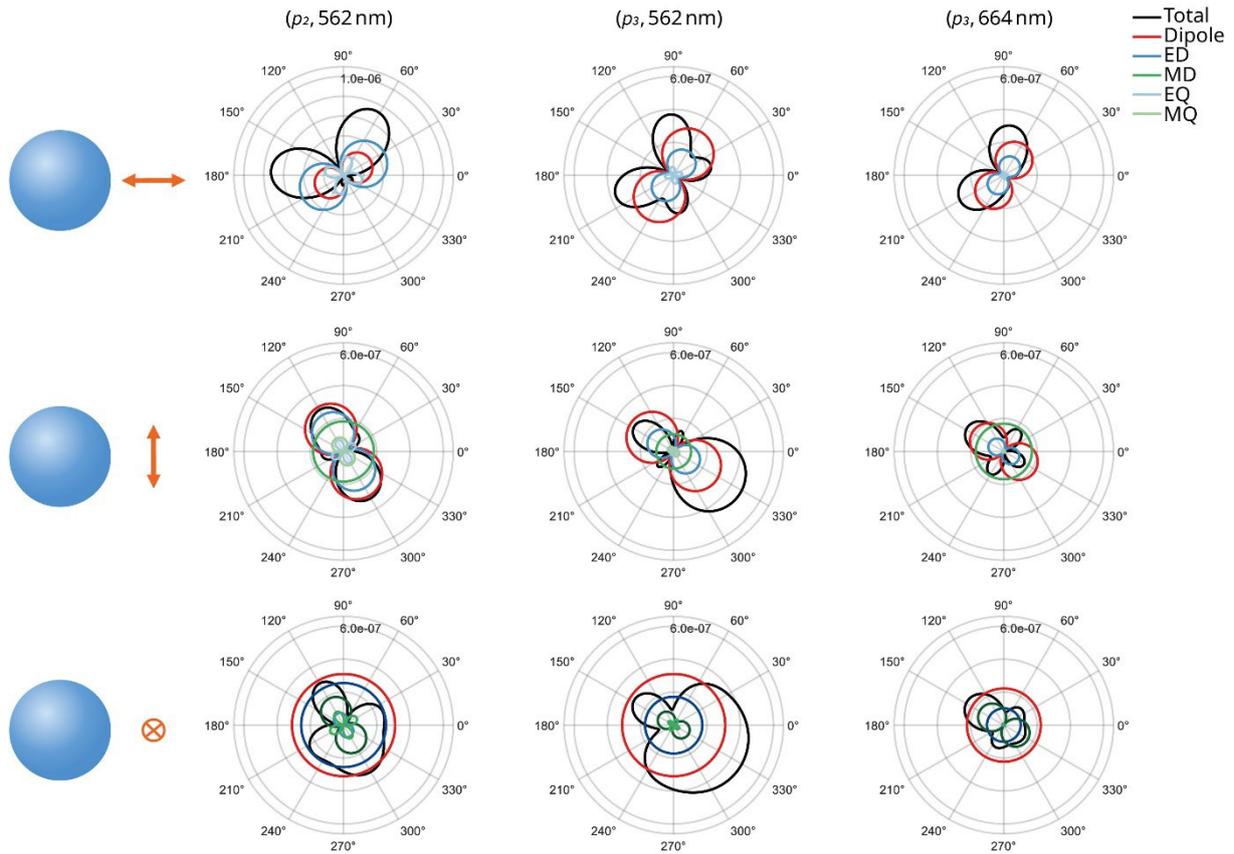

**Figure S8: Multipolar decomposition.** Multipolar decomposition of the far-field radiation patterns of a 170 nm SiNP excited by a dipolar source. Patterns are shown at different distances ($p_2$ and $p_3$) and emission wavelengths (562 and 664 nm) for the three possible orientations of the dipolar source (yellow arrow). The red line represents the dipole emitter, while ED, MD, EQ and MQ account for the induced multipoles in the SiNP.

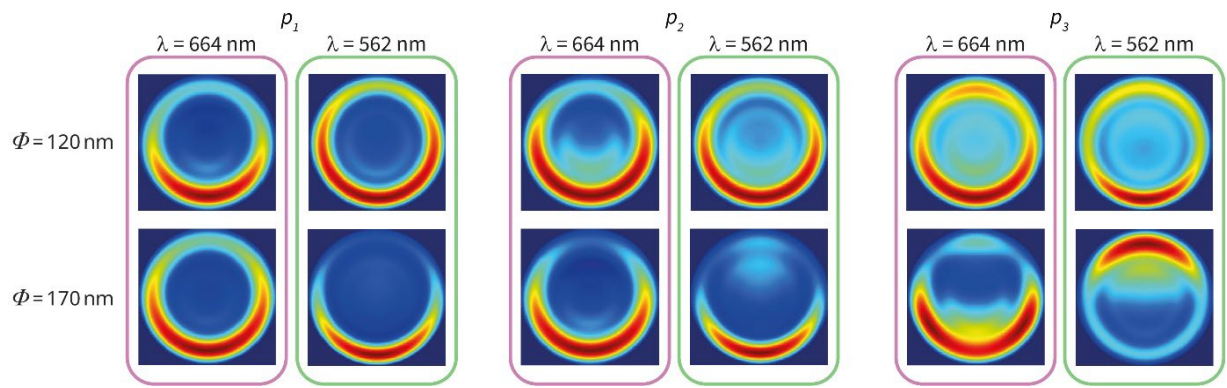

**Figure S9: BFP simulations.** Simulated BFP patterns for a SiNP of diameter 120 nm (top) and 170 nm (bottom) excited by a dipolar emitter (average orientation) located at positions $p_i$ from the SiNP surface. Patterns are shown for the emission wavelengths representing the two dyes used in our experiments. These patterns are used to compute F/B plots shown in Fig. 4.

Description of phasor diagrams in Figure 6D

For the case ($p_2$, 562 nm), as the longitudinal orientation suppresses the pure forward and backward directions, the angles 180º and 270º have been selected as forward and backward, respectively. As explained in the main text, the longitudinal orientation is characterized by electric resonances only, where a constructive interference between ED and EQ, together with the emitter radiation, gives a strong forward radiation (collector regime), and a destructive interference of the emitter with the ED, and the ED with the EQ, heavily suppresses the backward direction.

On the other hand, for the $p_3$ cases, forward and backward directions can be defined more easily as 150º and 300º, respectively. For a wavelength of 562 nm, we observe that the backward direction features a similarly constructive interference between the emitter radiation and the ED, with a partially constructive contribution from the MD. The constructive interference between emitted and ED does not occur at the forward direction, with a slightly-destructive contribution from the MD, leading to the overall backward, reflector regime.

Finally, the 664 nm wavelength at $p_3$ showcases a similar phasor diagram in the forward direction. In this case, however, the ED is mostly constructive with the emitter radiation, with the MD not having a particularly strong interference with the other modes. This situation is reversed in the backward direction, where it can be seen that the MD almost perfectly destroys the emitter radiation. This leads to a sort of generalized Kerker condition, where the backward radiation is heavily suppressed. Thus, the color-routing effect can be understood as a result of constructive interference between emitter and induced ED modes in one direction, with the MD displaying a destructive interference with the emitter in the other direction. The direction at which this happens depends on the wavelength, yielding the color-routing behavior.

| Name | Sequence |
|---|---|
| Biotin1 | ACAGGAAGATTGTCCCCCTTATTCACCCTCATTTGTTTC-Biotin |
| Biotin2 | GTTGATAGATATAAGCATAAGTATAGC-Biotin |
| Biotin3 | AGAGTACTCACGCTAACCTTTAATTGC-Biotin |
| Biotin4 | CACTAAAACACTCACGAACTAACACTAAAGT-Biotin |
| Biotin5 | TCACGACGTTGGGCGCTTTGGTAAAAC-Biotin |
| Biotin6 | CAGAGATAGCGATAGTGAATAACATAA-Biotin |
| ATTO542_P1 | AATTGAGTAATATCAGAAAATAAACAGCCATA-ATTO542 |
| ATTO647N_P1 | TACCCCGGTTAAAATTCCTTTGCCCGAACGTT-ATTO542 |
| ATTO542_P2 | ATTO542-CCCCCGCTGAGAGCCAGCAGGCCTGCAACAGTGCCACCCC |
| ATTO647N_P2 | ATTATCACCGGAAATGTTAGCAAACGTAGAA-ATTO647N |
| ATTO542_P3 | ATTATTACTTGGGAAGTTCATTACCCAAATCA-ATTO647N |
| ATTO647N_P3 | ATTO647N-CCCCCAGAGCGGGAGCTAACTTTCCTCGTTAGAATCCCC |
| ATTO542_P-2 | ATGCTGATGTTGGGTTTAGCTTAGATTAAGAC-ATTO542 |
| ATTO647N_P-2 | CCCCTACCGACAAAAGGTAATAAGAGAATATAAAGCCCCTT-ATTO647N |

**Table S1:** List of biotinylated and dye-modified staple sequences used in this project compared to the original design.[39] ATTO 647N and ATTO 542 staples present a single ATTO 647N or ATTO 542 fluorescent molecule for the alternative positions ($p_1$, $p_2$, $p_{-2}$ and $p_3$), and biotin staples allow binding to a neutravidin-functionalized glass surface.

| Name | Sequence |
|---|---|
| H.left 1(A8) | TACAAATTGCCAGTAAAGTAATTCTGTCCAGAAAAAAAA |
| H.left 2(A8) | TAAAGACTGTTACTTAGGCGCAGACGGTCAATAAAAAAAA |
| H.left 3(A8) | GCCACCCTTCGATAGCATAATCCTGATTGTTTAAAAAAAA |
| H.left 4(A8) | CACCAACCAAGTACAAGTACAGACCAGGCGCAAAAAAAA |
| H.left 5(A8) | CGACATTCCCAGCAAAATTATTTGCACGTAAAAAAAAAAA |
| H.left 6(A8) | CTTTTTTTTCATTTCAACAATAACGGATTCGAAAAAAAA |
| H.left 7(A8) | ACGGGTAATAAATTGTTGACCAACTTTGAAAGAAAAAAAA |
| H.left 8(A8) | ATTAATTATGAAACAATATACAGTAACAGTACAAAAAAAA |
| H.left 9(A8) | AATATTGACGTCACCGTGCGTAGATTTTCAGGAAAAAAAA |
| H.left 10(A8) | AACAGTAGCCAACATGACATGTTCAGCTAATGAAAAAAAA |
| H.left 11(A8) | AAGAACGCAAGCAAGCATAATATCCCATCCTAAAAAAAAA |
| H.left 12(A8) | CTTGCGGGGTATTAAAAAACCAATCAATAATCAAAAAAAA |
| H.left 13(A8) | AAATCAGATCATTACCATCAACAATAGATAAGAAAAAAAA |
| H.left 14(A8) | CATATGGTAGCAAGGTAATGGAAGGGTTAGAAAAAAAAA |
| H.left 15(A8) | CCATCTTTCGTTTTCAAACCACCAGAAGGAGCAAAAAAAA |
| H.left 16(A8) | GAGCCACCATCAAGTTTCCTGATTATCAGATGAAAAAAAA |
| H.left 17 (A18) | AAAAAAAAAAAAAAAAAAGGATTATACTTCTGAACCGGAAACGTCACCAA |
| H.left 18 (A18) | AAAAAAAAAAAAAAAAAAACCTACCATATCAAAATCACCAGTAGCACCA |
| H.left 19 (A18) | AAAAAAAAAAAAAAAAAAATTTACGAGCATGTAGCCAAGTACCGCACTCA |
| H.left 20 (A18) | AAAAAAAAAAAAAAAAAACATAAGGGAACCGAACGTCGAAATCCGCGACC |
| New Core 1 | GAAACGCAAAGACACCGCCAAAGATACCGAAG |
| New Core 2 | ATTTTGCACCCAGCTATTAGCGAAAGAATTAA |
| New Core 3 | AGCGAAAGACAGCATCAGGAAGTTTGTAGCAT |
| H.left 22(A18) | AAAAAAAAAAAAAAAAAAGCCGCCACTCATCAATAGCACCGTAATCAGT |
| H.left 23(A18) | AAAAAAAAAAAAAAAAAAATGGCAATCAGAACCCGCCTCCCTCAGGAGG |
| H.left 28(A8) | TGAGCGCTTAAGCCCATGGCATGATTAAGACTAAAAAAAA |
| H.left 29(A8) | CTGAACACATAGCAATAAACGCAATAATAACGAAAAAAAA |
| H.left 30(A8) | AAACAGGGAAGAAAAGGCCGAACAAAGTTACAAAAAAAA |
| H.left 31(A8) | TCCACAGAGGTGTATCGGATAAGTGCCGTCGAAAAAAAA |
| H.left 32(A8) | GTCACCAGCGCCACCCTTAGGATTAGCGGGGTAAAAAAAA |
| H.left 33(A8) | TAGGAACCCCACCCTCTATTAAGAGGCTGAGAAAAAAAAA |
| H.left 34(A8) | AATTAGAGAACCGATTGGCAACATATAAAAAAAAAAAAA |
| H.left 35(A8) | TTACCATTTTACCAGCACGGAATAAGTTTATTTAAAAAAAA |
| H.left 36(A8) | TGAAACCACAGAACCGAGCCACCACCCTCAGAAAAAAAAA |
| H.left 37(A8) | GCGACAGAACCGGAACACCACCAGAGCCGCCGAAAAAAAA |

**Table S2:** List of DNA modified staples, which were extended as NP capturing strands, used for the directional antennas compared to the original origami design.[39] Handles are staples extended with an PolyA (number of A indicated in parenthesis) to bind a single PolyT functionalized SiNP.

| i | $x_i$ (nm) | $\Phi$ = 120 nm $|d_i|$ (nm) | $\Phi$ = 170 nm $|d_i|$ (nm) |
|---|---|---|---|
| -2 | -50 | 20 | 15 |
| 1 | -25 | 8 | 6 |
| 2 | 50 | 20 | 15 |
| 3 | 133 | 85 | 75 |

**Table S3:** List of fluorophore positions in the DNA origami (taking the SiNP as $x = 0$) and estimated distance $d_i$ to the SiNP surface depending on its diameter.